\documentclass[12pt]{article}
\usepackage{geometry} 
\usepackage{algorithm}
\usepackage{algorithmic}
\usepackage{amsmath}
\usepackage{enumerate}
\usepackage{natbib}
\usepackage{amsthm}

\theoremstyle{definition}

\usepackage{array}
\newcolumntype{C}[1]{>{\centering\arraybackslash}p{#1}}
\newcolumntype{C}[1]{>{\centering\let\newline\\\arraybackslash\hspace{0pt}}m{#1}}

\usepackage{subcaption}
\usepackage{graphicx,psfrag,epsf,color}
\usepackage{float,booktabs}
\usepackage{multirow}

\usepackage{hyperref}
\graphicspath{{figures/color/}} 

\newcommand{\blind}{1}

\date{} 

\begin{document}

\def\spacingset#1{\renewcommand{\baselinestretch}%
{#1}\small\normalsize} \spacingset{1}

\if1\blind
{
  \title{\bf Robust Clustering with Subpopulation-specific Deviations}
  \author{Briana J.K. Stephenson\thanks{Briana J.K. Stephenson is Postdoctoral Research Associate, University of North Carolina at Chapel Hill, Chapel Hill, NC 27599 (e-mail: \url{bjks@unc.edu}). Amy H. Herring is Sara \& Charles Ayres Professor of Statistical Science, Duke University, Durham, NC 27709 (e-mail: \url{amy.herring@duke.edu}). Andrew Olshan is Barbara Hulka Distinguished Professor, Department of Epidemiology, University of North Carolina at Chapel Hill, Chapel Hill, NC 27599.
    The authors gratefully acknowledge this work was supported in part through cooperative agreements from the Centers for Disease Control and Prevention to the centers participating in the National Birth Defects Prevention Study and by the National Institutes of Health (R01ES020619; R01ES027498; U50CCU422096; 5U01DD001036; PA 96043; PA 02081; FOA DD09-001; NIEHS T32ES007018). Dietary and nutrition information was made possible by the University of North Carolina Epidemiology Core (Grant No: DK56350). We thank Willem van den Boom for assistance in data replication. Finally, we thank Louise Ryan and David Dunson for helpful comments on earlier drafts of this work.}\hspace{.2cm}\\
    Department of Biostatistics, University of North Carolina at Chapel Hill\\
    Amy H. Herring\\
   Department of Statistical Science, Duke University \\
     Andrew Olshan \\
    Department of Epidemiology, University of North Carolina at Chapel Hill\\
    The National Birth Defects Prevention Study}
  \maketitle
} \fi

\if0\blind
{
  \bigskip
  \bigskip
  \bigskip
  \begin{center}
    {\LARGE\bf Robust Clustering with Subpopulation-specific Deviations}
\end{center}
  \medskip
} \fi

\bigskip

\begin{abstract}

The National Birth Defects Prevention Study (NBDPS) is a case-control study of birth defects  conducted across 10 U.S. states. Researchers are interested in characterizing the etiologic role of maternal diet, collected using a food frequency questionnaire. Because diet is multi-dimensional, dimension reduction methods such as cluster analysis are often used to summarize dietary patterns. In a large, heterogeneous population, traditional clustering methods, such as latent class analysis, used to estimate dietary patterns can produce a large number of clusters due to a variety of factors, including study size and regional diversity. These factors result in a loss of interpretability of patterns that may differ due to minor consumption changes. 

Based on adaptation of the local partition process, we propose a new method, Robust Profile Clustering, to handle these data complexities. Here, participants may be clustered at two levels: (1) globally, where women are assigned to an overall population-level cluster via an overfitted finite mixture model, and (2) locally, where regional variations in diet are accommodated via a beta-Bernoulli process dependent on subpopulation differences. We use our method to analyze the NBDPS data, deriving pre-pregnancy dietary patterns for women in the NBDPS while accounting for regional variability.
\end{abstract}

\noindent%
\vfill

\newpage
\spacingset{1.45} 


\section{Introduction}
\subsection{Multivariate Categorical Dietary Data}
Food frequency questionnaires (FFQ) are often used to measure an individual's dietary intake over a period of time. The standard FFQ queries consumption/intake levels for over 100 foods and beverages \citep{subar2001comparative}. Some researchers focus on individual foods or nutrients, but foods are not consumed in isolation, and many nutritionists argue that a more holistic approach is needed \citep{motulsky1989diet}. When data include a large number of exposures or, in this case, food items of an FFQ, data reduction techniques such as factor analysis, latent class analysis, or other clustering approaches are often used \citep{kant2004dietary, venkaiah2011application, sotres2010latent, keshteli2015patterns}.  

Clustering methods generally assume participants within each cluster share dietary habits and aim to maximize differences across groups. At times, these methods may oversimplify dietary behaviors, and it can be difficult to determine when dimension reduction is generalizable across different populations. The generalizability issue poses a concern for large heterogeneous populations. Subjects from these populations may often share a combination of  behaviors that could be general to an overall population, but also specific to a subject's subpopulation, defined as any group indicated by a categorical covariate (e.g. state residence, ethnicity, SES, etc.). For example in the United States, if a subpopulation was defined by a subject's state of residence, the foods consumed to characterize ``American'' diets would look different, due to regional differences. A healthy diet may incorporate an increased consumption of regional foods indigenous to a specific state (e.g. more avocados in Texas). Reconciling these regional differences with a single overall clustering method presents a loss of granularity.

On the other hand, creating separate models for each subpopulation can greatly diminish statistical power, and can lead to misleading characterizations of diets when generalizing across the entire population sample.  Individuals that are classified as having a ``healthy'' diet in North Carolina, or a subpopulation where poor eating behaviors are prevalent, may be classified as having an ``unhealthy'' diet in Massachusetts, where a more `health-conscious' subpopulation is prevalent. The differences found within these regional patterns are crucial for the improvement of national dietary recommendations that can accommodate heterogeneity of dietary behaviors. 

\subsection{Standard Clustering Methods}
The latent class model, introduced by \citet{lazarsfeld1968latent}, is the most common clustering method for handling multivariate categorical response data. Because the number of clusters is typically unknown, models with a varying number of clusters are fit. The best number of clusters is often chosen via likelihood ratio tests, Bayesian Information Criteria, Akaike Information Criteria, or the Lo-Mendell-Rubin test \citep{nylund2007deciding}. In practice, these criteria tend to be ``greedy" and select solutions with a large number of clusters. To avoid the challenges of a large number of clusters, in practice researchers will often add interpretability as a criterion for selecting the number of clusters and look for the best solution with a ``manageable'' number \citep{ford2010poly, silverwood2011lca}. With no separation across subpopulations, this restriction on the number of overall classes may mask any localized dietary behaviors that could be pronounced at a subpopulation level.

Nonparametric Bayesian methods, e.g., the Dirichlet process or overfitted finite mixture models, allow the number of clusters represented in a sample to grow as dimensions (sample size, number of variables) increase \citep{figueiredo2002unsupervised, zhang2004probabilistic, teh2006hierarchical, miller2018mixture, rousseau2011asymptotic}. In heterogeneous populations, a largely prevalent subpopulation may have its behaviors reflected in one of the overall clusters, while smaller subpopulation behaviors may still remain hidden in one of those same clusters. Further, while flexibly convenient, these models tend to overestimate the true number of clusters, permitting, at times, nonexistent clusters to appear \citep{miller2013simple}. Outliers are often assigned to singleton clusters, which may measure lack of fit in the model more than a new pattern. 

In multi-site studies, often a hierarchical or nested approach is used to accommodate any potential differences amongst subpopulations. The hierarchical Dirichlet process assumes common clusters across groups \citep{teh2012hierarchical}. Nested approaches cluster subjects within a subpopulation and borrow information across subpopulations that share similar behaviors \citep{rodriguez2012nested, hu2018dirichlet}. While useful in many applications, these techniques contain drawbacks. The number of nonempty clusters derived is highly sensitive to the selection of tuning parameters or hyperparameters. Often unrealistic, strong priors on these parameters are necessary to enforce sparsity and ensure subjects aggregate to a reasonable number of clusters, as interpretability once again becomes an issue. 

What strains these clustering methods is the assumption of global clustering, where subjects belonging to the same cluster will exhibit the same expected set of responses for all variables included in the set. This is where subpopulation granularity is lost. Differences may exist for a subset of variables within that subpopulation. Local partition and hybrid Dirichlet process mixture models break this global clustering assumption by apply a two-tiered clustering scheme at a global and local level \citep{dunson2009lpp, petrone2009hybrid}. 

The hybrid Dirichlet process mixture model makes local cluster assignments to each individual variable and then clusters globally based on shared similarities of the local clusters with other subjects using a copula construction. The local clustering is considered a smaller subset of the overall population, but it is not specific to any identifiable subpopulation. Additionally, the hybrid Dirichlet prior is limited to continuous data. FFQ data are considered semi-quantitative, as quantity of consumption is collected based on choices from several standardized portion sizes. The frequency of consumption is collected in ordinal group levels (e.g. X times per day, daily, weekly, monthly) \citep{subar2001comparative}. Given this data structure, the copula construction used in the hybrid DP is unable to generate unique solutions to discriminate global clusters when handling discrete data \citep{smith2012estimation}. A hierarchical hybid Dirichlet could be proposed that would deal with discrete data and allow another level in the dependence structure. Such a model has not yet been proposed in the literature, to our knowledge.

Instead of clustering every variable individually, the local partition model allows an entire subset of variables to be partitioned to a local or global clustering system. Here, subjects cluster with other subjects who behave similarly for most or some of the variables analyzed. It is useful in characterizing the global cluster patterns because variables that do not provide much information to the overall population patterns can be assigned to a local cluster. As with the hybrid DP, the local level is not specific to any identifiable subpopulation. Those variables that are considered noise at the global level could be valuable at a subpopulation level. In order to identify which items are important in a general population setting and which items are important in a subpopulation setting, a statistically principled method is needed to identify and discriminate between the two levels of patterns, while still preserving a level of interpretability.

We organize this article as follows. Section 2 introduces the RPC framework. Section 3 explores RPC functionality and performance against other methods using a simulation study. Section 4 presents a comprehensive analysis of the NBDPS data and describes insights provided by the new methodology. We conclude with a short discussion in Section 5.

\section{Robust Profile Clustering}
In this section, we propose a novel class of Robust Profile Clustering (RPC) processes, which are designed to produce a robust set of ``global'' clusters summarizing the overall nutritional profile of an individual. Clustering does not follow the typical approach of restricting all of the measurements from individual $i$ to have the same cluster membership. Instead, local deviations from the global profile are allowed and explored within each identifiable subpopulation. We define this style as robust due to the model's ability to derive a global cluster that is robust to individual subpopulations that do not follow the global patterns everywhere. This robustness prevents RPC from introducing extraneous clusters to fit small deviations from global profiles within subpopulations.  


Our data are nested within subpopulations. This allows local deviations to have a subpopulation-specific form. Introducing some notation, we let $i=1,\ldots,n$ index individuals in a study, $s_i \in \{1,\ldots,S\}$ index the known subpopulation (essentially a categorical covariate) of individual $i$, and $C_i$ index the (unknown) global profile membership of subject $i$. Additionally, each subject has a multivariate data vector, $\mathbf{y}_i=(y_{i1},\ldots, y_{ip})'$. Individual $i$ may not follow her global cluster allocation for all elements of this multivariate vector but may deviate for some as needed. We let $G_{ij} = 1$ if item $j$ is attributed to global cluster $C_i$ for individual $i$ and $G_{ij}=0$ otherwise. We let $L_{ij}$ denote the local cluster allocation conditionally on $G_{ij}=0$ and $s_i=s$.

An RPC process is then induced through probability models containing 3 components: (\textit{i}) the global clustering, $C_i$, (\textit{ii}) the variable deviation indicator, $G_{ij}$, and (\textit{iii}) the local clustering membership, $L_{ij}$. There are a wide variety of choices that can be used for (\textit{i})-(\textit{iii}), and to put in general form we let

\begin{equation}\label{rpc0}
\begin{aligned}
Pr(C_i=h) &= \pi_{h}, \\ Pr(G_{ij}=1|s_i=s) &= \nu_{j}^{(s)}, \\ Pr(L_{ij}=l | s_i=s) &= \lambda_{l}^{(s)}
\end{aligned}
\end{equation}

The data ${\bf y}_i=(y_{i1},\ldots,y_{ip})'$ are assumed to be drawn independently from a set of global density parameters $\Theta_{0\cdot C_i,\cdot} = \{ \theta_{0jC_i, \cdot} \}_{j=1}^p$ or local density parameters $\Theta_{1\cdot L_{ij},\cdot}^{(s)}=\{\theta_{1jL_{ij}, \cdot}^{(s)}\}_{j=1}^p$. For example, in a multivariate categorical case, we can use an equal number of categories $d$ for simplicity in exposition, where $\theta_{0jC_i,\cdot} = (\theta_{0jC_i,1},\ldots, \theta_{0jC_i,d})'$ and $\theta_{1jL_{ij},\cdot} = (\theta^{(s)}_{1jL_{ij},1},\ldots, \theta^{(s)}_{1jL_{ij},d})'$. An extension to a variable number of categories is easily accommodated. We then let 

\begin{equation}
y_{ij} \sim \begin{cases} \text{Mult}(\{1,\ldots,d\}, \theta_{0jC_i,\cdot}) & \text{if } G_{ij}=1 \\ \text{Mult}(\{1,\ldots,d\}, \theta^{(s)}_{1jL_{ij},\cdot}) & \text{if } G_{ij}=0, s_i=s,
\end{cases}
\end{equation}
where the cluster- and food item-specific probability vectors $\{ \theta_{0jC_i,\cdot}, \theta^{(s)}_{1jL_{ij},\cdot} \}\overset{iid}{\sim}\text{Dirichlet}(1,\ldots,1)$ \emph{a priori} for $j=1,\ldots,p$, $C_i=1,2,\ldots$, $L_{ij} = 1,2, \ldots$, $s=1,2,\ldots,S$.  Although $y_{ij}$ and $y_{ij'}$ are conditionally independent given $\boldsymbol{\theta}$, dependence is induced in marginalizing out the global cluster index $C_i$, as shown in expression \ref{rpclik} below.

We assume binary deviation vectors, $G_{i\cdot} = (G_{i1},\ldots, G_{ip})$, for all $i \in (1,\ldots, n)$, are independent and identically distributed given $s_i=s$ with deviation probability $\nu_{j}^{(s)}$. Given the binary structure of a variable's deviation to one of two levels, we model each subpopulation with a Beta-Bernoulli process to exploit its convenient conjugacy properties. 

\begin{equation}
G_{ij} \sim \text{Bern}(\nu_j^{(s)}), \qquad \nu_j^{(s)} \sim \text{Be}(1,\beta^{(s)}), \qquad \beta^{(s)} \sim \text{Ga}(a,b).
\end{equation}

The hyperparameters $(a,b)$ control the overall weight given to each local component (deviated food item) of its respective subpopulation. We let $a=b=1$ as a default to place equal probability \emph{a priori} on the global and local components, while allowing substantial uncertainty.

For the global clustering process, we assume an overfitted finite mixture model \citep{van2015overfitting}, which greatly simplifies computation relative to the LPP. Let $K$ be a conservative upper bound on the number of clusters (say, K=50). Then we have
\begin{equation}
\begin{aligned}
Pr(C_i=h)&= \pi_h \\
\pi_\cdot &= (\pi_1,\ldots, \pi_K)' \sim \text{Dir}\left(\frac{1}{K}, \ldots, \frac{1}{K}\right).
 \end{aligned}
 \end{equation}
 
For the local clustering process, we use a parallel formulation, letting
 \begin{equation}
 \begin{aligned}
Pr(L_{ij}=l \mid s_i=s)&= \lambda_l^{(s)} \\
\lambda_\cdot^{(s)}&= (\lambda_1^{(s)},\ldots, \lambda_K^{(s)}) \sim \text{Dir}\left(\frac{1}{K}, \ldots, \frac{1}{K}\right).
 \end{aligned}
 \end{equation}

The induced subject-specific likelihood ${\bf y}_i$ conditionally on $G_{i\cdot}=\{G_{ij}\}_{j=1}^p$,  $\Theta_{0\cdot C_i,\cdot} = \{ \theta_{0jC_i, \cdot} \}_{j=1}^p$, $\Theta_{1\cdot L_{ij},\cdot}^{(s)}=\{\theta_{1jL_{ij}, \cdot}^{(s)}\}_{j=1}^p$, but marginalizing out $C_i$ and $L_{i\cdot}=\{L_{ij}\}_{j=1}^p$ is given by

\begin{equation}\label{rpclik}
f(\mathbf{y}_i|-) = \left[\sum_{h=1}^{K} \pi_h \prod_{j:G_{ij}=1}^p\prod_{r=1}^{d} \theta_{0jh,r}^{\mathbf{1}(y_{ij}=r)}\right] \prod_{j:G_{ij}=0}^p\left[\sum_{l=1}^{K} \lambda_{l}^{(s)} \prod_{r=1}^{d} (\theta^{(s)}_{1jl,r})^{\mathbf{1}(y_{ij}=r)}\right].
\end{equation}

\section{Simulation Study}

We use this section to explore the performance of existing methods with the newly proposed RPC via a simulation study. We observed model performances under seven scenarios: (1) strictly global, where data contain only true underlying global clusters, and no local clusters, (2) strictly local, where data contain only true underlying local clusters, but no global clusters, (3) hybrid case, where each subpopulation contained an increasing proportion of deviated variables to a local profile, (4) null case where the data contained no true underlying clusters at the global or local level, (5) a combination of scenarios (1)-(3) to mimic a large, heterogeneous population as expected in the NBDPS dietary dataset, (6) various settings violating our Beta-Bernoulli assumption, and (7) an exploration of the model under a continuous data setting. 
A total of 500 datasets were generated for each scenario across a pre-identified finite set of subpopulations $S$. Each dataset described a total of p=50 variables that each contain $d=4$ categorical response variables. Continuous variables were randomly drawn from a normal distribution, according to its predefined cluster assignment. Variables were probabilistically allocated to global or local levels by pre-specified fixed values of $\nu_j^{(s)}$. Exact specifications for each simulation case are detailed in Appendix A.

\subsection{MCMC computation}
Discrete data were analyzed under 5 different models: (1) traditional LCA model with four classes (tLCA) \citep{lazarsfeld1968latent}, (2) Dirichlet process mixture model (DPM) \citep{dunson2009nonparametric}, (3)  over-fitted finite mixture model of $K=50$ classes (oFMM) \citep{rousseau2011asymptotic}, (4) local partition process model ($LPP_2$) \citep{dunson2009lpp}, and (5) the newly proposed robust profile clustering model (RPC). Continuous data were analyzed under a sparse finite Gaussian mixture (GMM) of $K=50$ classes and RPC models. While models (1)-(4) ignore any subpopulation effect, we include them to demonstrate how the RPC properties compare to the methods currently available. 

Estimation was performed using a Gibbs sampler of 20,000 runs after a 5,000 burn-in. Posterior medians of all model parameters were calculated from the MCMC output. To encourage mixing, label switching moves were imposed to allow better exploration of the parameter space. The random permutation sampler was applied for finite mixture model cases (tLCA, oFMM, GMM, RPC) \citep{fruhwirth2001markov}. The DPM and $LPP_2$, which involve a Dirichlet process mixture model, had two label switching moves imposed to favor swapping of both equal and unequal size clusters \citep{papaspiliopoulos2008retrospective}. 

Mixing efficiency and convergence were evaluated using trace plots of model parameters and randomly selected variables. Parameters were relabelled using a similarity matrix from the MCMC output that contained the pairwise posterior probabilities of two subjects being clustered together in a given iteration. Hierarchical clustering was then performed on the similarity matrix, using the complete linkage approach \citep{krebs1989ecological,medvedovic2002bayesian}. A nonempty cluster was identified as any cluster with a posterior probability weight greater than 0.01. 

Dirichlet hyperpriors of the finite mixture component weights found in the tLCA, oFMM, GMM, RPC were preset to $1/K_{\max}$, where $K_{\max}$ is the preset maximum number of clusters allowed in the model. The concentration hyperparameter of models containing a Dirichlet process was preset to 1. Parameters estimated from all existing methods were sampled in accordance with their algorithms found in their respective literature \citep{nylund2007deciding, dunson2009nonparametric, rousseau2011asymptotic, dunson2009lpp}. 

In the continuous case, normal gamma priors were generated based on the range of input data, such that the prior mean for each cluster was drawn from a normal distribution with mean 0 and variance $\tau = 10$. Cluster specific variance at both the global and local level were drawn from a Gamma distribution with shape a = 0.1 and scale b = 10. 

Profile patterns from each respective model were derived using the posterior median of cluster density parameter estimates. Each cluster density contained a vector of posterior probabilities of a subject responding at a given level within that cluster. The response level containing the maximum probability of each variable was designated as the modal cluster pattern response. The modal cluster responses were used to identify any potential redundant clusters present. Sensitivity was measured as the proportion of subjects belonging to the same pre-assigned cluster were assigned to the same cluster by the model. Specificity was defined as the proportion of subjects who were not pre-assigned together remained not assigned together. Density concordance was measured by comparing how well each model-derived cluster density compared to the true cluster density. This was calculated by calculating the mean squared error $(MSE_\theta)$ of the true density parameters, $\{\theta_{0jC_i}\}_(j=1)^p$, to corresponding model cluster density parameters, $\{\hat{\theta}_{0jC_i}\}_(j=1)^p$. 
To evaluate RPC's discriminating feature of determining if a variable is global ($\hat{\nu}_j^{(s)} \to 1$) or local ($\hat{\nu}_j^{(s)} \to 0$), total MSE was calculated for each dataset, such that $MSE_\nu = \frac{\sum_s \sum_j (\hat{\nu}_j^{(s)} - \nu_j^{(s)})}{S \cdot p}$, where $\hat{\nu}_j^{(s)}$ is the posterior median estimate of $\nu_j^{(s)}$. Variables predicted as global will have higher probabilities approaching 1 and variables predicted as local will have lower probabilities approaching 0. A small $MSE_\nu$ implies a strong discrimination of the RPC model across all $p$ variables. 

\subsection{Results}\label{sec:simresults}
All models were run using MATLAB (version 2017a). Examination of trace plots indicated good mixing and convergence of model parameters.  The median number of nonempty clusters was collected across all 500 simulated sets for each case. The median summaries across all 500 simulated sets is summarized in Table \ref{tab:comp}. The number of nonempty global clusters derived in each model is found where  $K_0 =\sum_k \mathbf{1}(\hat{\pi}_k > 0.01$), where $\hat{\pi_k}$ is the posterior median component weight of cluster $k$. An issue with redundant clustering was evident in the RPC and on occasion the LPP and GMM model. This is a common consequence when using sparse mixture models \citep{malsiner2016model}. These models required an additional post-processing step to remove redundant clusters. The final set of unique nonempty clusters can be found by subtracting the number of redundant clusters from $K_0$. Global/Local discrimination results were illustrated in the last column, $MSE_\nu$. 
\begin{table}
\centering
\resizebox{!}{0.5\textheight}{
\begin{tabular}{|c|c|c|c|c|c|}
\toprule
Case & Model & $K_0$ (IQR) & Redundant Clusters & Concordance $MSE_\theta$ & $MSE_\nu$ \\
\hline
\multirow{5}{*}{Global} & LCA & 3 (0) & 0 & 0.09 & NA\\
	& DPM & 3 (0) & 0 & $< 0.01$ & NA\\
	& OFM & 3 (0) & 0 & 0.08 & NA \\
	& LPP & 3 (0) & 0 & $< 0.01$ & NA \\
	& RPC & 5 (0) & 2 & 0.01 & 0.07 \\
\hline
\multirow{5}{*}{Local} & LCA & 4 (0) & 0 & 0.07 & NA\\
	& DPM & 8 (0) & 0 & $< 0.01$ & NA\\
	& OFM & 8 (0) & 0 & 0.08 & NA\\
	& LPP & 7 (1) & 0 & $< 0.01$  & NA \\
	& RPC & 2 (0)* & 0 & 0.01 & $< 0.01$\\
\hline
\multirow{5}{*}{Hybrid} & LCA & 4 (0) & 0 & 0.06 & NA \\
	& DPM & 17 (2)  & 0 & $<0.01$ & NA\\
	& OFM & 15 (1) & 0 & 0.08 & NA\\
	& LPP & 8 (1) & 0 & 0.01 & NA \\
	& RPC & 6 (1) & 3 & 0.01 & 0.03 \\
\hline
\multirow{5}{*}{Null} & LCA & 1 (0) & 0& NA & NA\\
	& DPM & 1 (1) & 0 & NA & NA\\
	& OFM & 1 (1) & 0 & NA & NA\\
	& LPP & 47 (3) & 0 & NA & NA \\
	& RPC & 1 (0) & 0 & NA & NA \\
\hline
\multirow{5}{*}{Mock NBDPS} & LCA & 3 (1) & 0 & 0.57 & NA \\
	& DPM & 18 (2)  & 2 & $<0.01$ & NA\\
	& OFM & 17 (1) & 0 & 0.08 & NA\\
	& LPP & 12 (1) & 1 & $<0.01$ & NA \\
	& RPC & 8 (1) & 4 & 0.01 & 0.03 \\
\hline
Beta-Bernoulli & RPC & 5 (0) & 2 & $<0.01$ & 0.03\\
\hline
\multirow{6}{*}{Continuous} & Separate: GMM & 3 (0) & 1 & $< 0.01**$ & NA\\
	& Separate: RPC & 3 (0) & 1 & $<0.01**$ & 0.06\\
	& Partial: GMM & 6 (0) & 4 & $<0.01**$ & NA\\
	& Partial: RPC & 6 (2)& 4 & $<0.01**$ & 0.03\\
	& Overlap: GMM & 0 (0) & 0 & NA & NA\\
	& Overlap: RPC & 34 (4) & 10 & 0.06** & 0.07\\
\bottomrule
\end{tabular}
}
\caption{Table summary of simulated cases (1)-(7) for 500 respectively simulated datasets. $K_0$: median number of nonempty global clusters with a weight larger than 0.01. Concordance: median mean square error of true cluster density parameters and corresponding model cluster density parameters. 
*Local profile patterns were not reflected at the global level
**For the continuous case, concordance was only compared for variables that were pre-allocated global across both subpopulations. \label{tab:comp}}
\end{table}

DPM, OFM, LPP, and RPC models performed well under the completely global case (1) and completely local case (2). The true cluster modes were identified with no additional clustering. With the exception of the RPC, all of the expected cluster patterns in the model were reflected at the global level in both cases. In the completely local (case 2) setting, patterns were masked in the LCA model due to the forced model restriction to $K=4$ classes. In both of these cases, the RPC performed well in being able to recognize that all of the variables in the completely global case (1) should be assigned at the global level with an overall mean $\nu_j^{(s)} = 0.75 \pm 0.001$ across all variables $j \in (1,\ldots, p)$ and subpopulations $s \in (1,\ldots, S)$. As a result, all of the patterns in case 1 were reflected at the global level. Similarly, in the completely local setting (case 2), the RPC was able to recognize that all of the variables should be local with an overall mean $\nu_j^{(s)} = 0.03 \pm .0002$ across all variables and subpopulations. Consequently, the patterns in the completely local case were reflected at the local level. With all of the variables in the global setting (case 1) and local setting (case 2) favoring one level in the RPC model, few if any clusters were identifiable at the non-favored level. For example, two global profiles were derived in the local setting (case 2). However, the posterior probability corresponding to the modal response level of these profiles was 0.25, indicating no strong modal response for any variables or profiles at the global level. 

With no true pattern evident in the null case (4), all of the models collapsed participants into a single cluster, with the exception of $LPP_2$. The $LPP_2$  identified an overwhelming number of small clusters (median = 47). The RPC model indicated a strong tendency to the single global cluster with $\nu_j^{(s)} > 0.99$ across all variables and subpopulations. As a result, at the local level no identifiable clusters were formed within each subpopulation. These results are consistent in reflecting no true pattern in the population set. 

All of the models were able to identify the expected modes of each global profile pattern in case (3). However, LCA, DPM, OFM, $LPP_2$ models had a decreased sensitivity (57\%) as additional clusters were derived to accommodate the hybridization of global and local variables. This resulted in subjects who originally belonged to the same cluster being dispersed across several similar, but not identical clusters. The RPC was the only model that had a high sensitivity (99\%) and specificity (99\%).  It derived the true global patterns without additional clustering. At the local level, the model's imposed sparsity resulted in a single local cluster per subpopulation. These singleton clusters derived reflected a modal split between the two expected local profile modes. For example, if local variable $j$, had a response level of 4 for half the subpopulation and a response level of 3 for the remaining half of the subpopulation, the single local cluster would reflect an approximate probability of 0.5 at response level 3 and also at response level 4. 

\begin{figure}[H]
  \centering
    \includegraphics[width=0.8\textwidth]{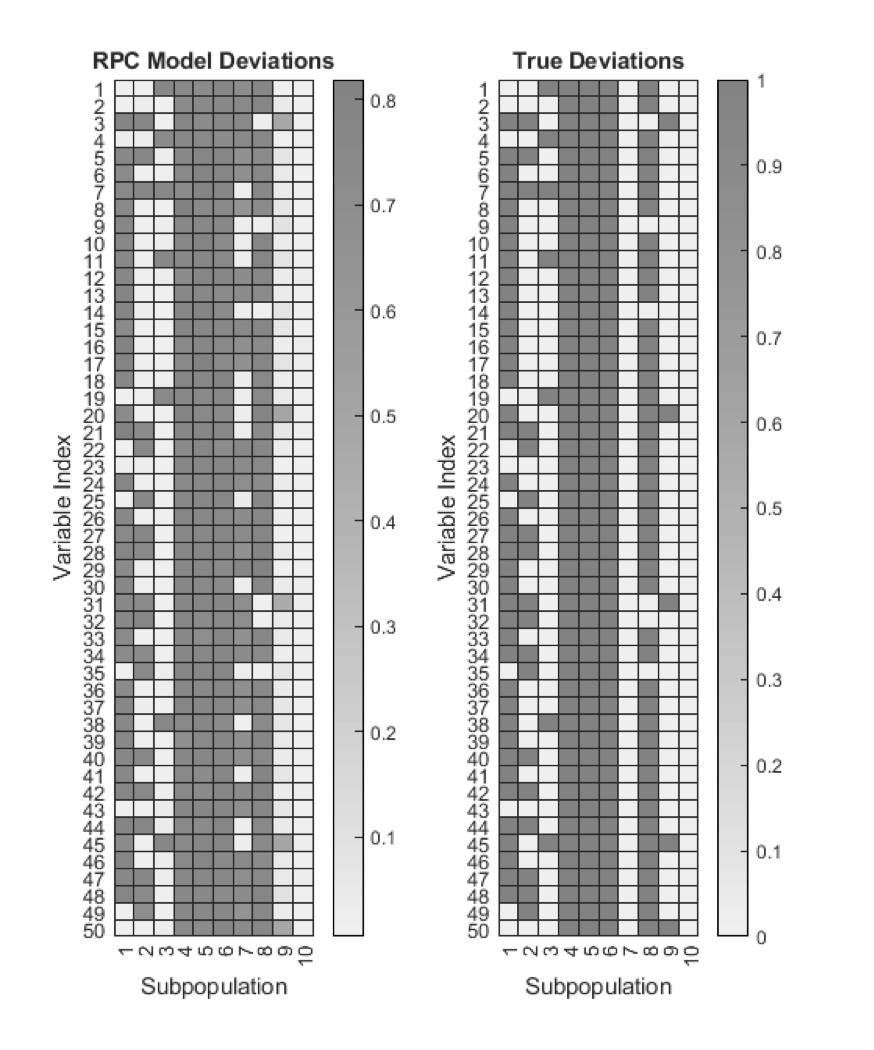}
      \caption{Simulation example comparing probability of a variable being allocated to a global cluster set (gray,$G_{ij}=1$) and local cluster set (white, $G_{ij}=0$). Side-by-side figure compares RPC-derived results (left) with true deviation results (right).}\label{fig:RPCnu}
\end{figure}

These results were consistent in the expanded mock NBDPS case (5). Due to subpopulations containing an increasing proportion of locally allocated variables, all of the models with the exception of RPC derived several extraneous clusters. The RPC had only one extraneous cluster, which was due to a subpopulation that reflected a completely local profile. The discrimination feature of the RPC model performed well in both the hybrid case (3) and mock NBDPS case (5) easily identifying which variables should be global or local ($\overline{MSE}_\nu = 0.03$, Figure \ref{fig:RPCnu}).

The $\{\beta^{(s)}\}_{s=1}^S$ parameter in the RPC model illustrates the weight of each subpopulation to the model's global or localized components. Lower values indicate a strong representation of global components. For example, in global case (1), the $\beta^{(s)}$ for each of the subpopulations were all less than 1. Higher values indicate a stronger representation of local components, as illustrated in local case (2), where $\beta^{(s)}=19$ for each of the subpopulations.  

Consistent with prior simulation cases, RPC was still able to preserve a strong level of specificity (99\%) and sensitivity (99\%) to the global and local profile patterns. The $\overline{MSE}_\nu$ for each subpopulation were consistent with RPC results of prior simulation cases. This consistency implies the model's robustness to violations of the beta-Bernoulli assumption.

In the continuous case (7), redundancy in cluster profiles increased as the standard deviation increased for both RPC and GMM. The RPC performed well in highlighting the global/local discrimination feature with an $\overline{MSE}_\nu \le 0.07$ in all three sub-cases. This feature led to a split in pattern identification. The expected global profile information was consistent with the expected global mean only when it was assigned to the global level. Likewise, expected local information was consistent with the expected local mean only when it was assigned to the local level. For example, all variables assigned to global profile 1 had a mean near 2.0, as expected. If variable 6 within subpopulation 1 was expected to deviate to the local level, the local profile densities at variable 6 reflected an expected mean of 5 for local profile 1 and -5 for local profile 2, as expected. When concordance was compared with only variables that were global across all subpopulations, both GLM and RPC were able to identify the parameter values. 

\section{Analysis of National Birth Defects Prevention Study Data}
\subsection{Multivariate Categorical Dietary Data}
The National Birth Defects Prevention Study is an ongoing multi-state population-based, case-control study of birth defects in the United States \citep{yoon2001national}. Infants were identified via birth defect surveillance in Arkansas, California, Iowa, Massachusetts, New Jersey, New York, Texas, Georgia, North Carolina, and Utah. We focus this analysis on dietary habits of control participants. Participants for this study included mothers with expected due dates from 1997 to 2009, totaling 9747 controls. Controls were defined as any live-born infant without any birth defects and were randomly selected from birth certificates or hospital records. Following \citet{sotres2013maternal}, subjects were excluded who had multiple births, a prior history of birth defects, preexisting diabetes, or folate antagonist medication use from three months before pregnancy to the end of pregnancy, and unexplained. Mothers with daily energy intake below the 2.5th and above the 97.5th percentiles were also excluded to prevent inclusion of unlikely intake data. After exclusion criteria were applied, a total of 9010 controls were included for this analysis. 

Food consumption was measured in grams per day and calculated by multiplying frequency of consumption by the standard portion size for each food item as outlined in \citet{sotres2013maternal}. Because respondents are prone to over-estimate or under-estimate total intake \citep{t1991validation,haraldsdottir1993minimizing}, percentiles were computed by dividing an individual food consumed over the total foods consumed, using grams per day as the consumption metric. The distribution of food items showed a spike at zero, which is well known in the literature \citep{kipnis2009modeling,zhang2011new}. Keeping with common dietary analysis practices, these percentiles were aggregated into four relative consumption levels: no consumption, low consumption (0-33\% consumed), medium consumption (33-66\% consumed), and high consumption (66-100\% consumed). A total of $p=63$ food items were included in the study with four consumption levels $(d=4)$ fit using a categorical distribution. 


For our RPC model, we define $\theta_{0jk,r}$ as the probability of a subject having a consumption level of $r$ from food item $j$, given allocation to global dietary profile $C_i = k$. Similarly, we define $\theta^{(s)}_{1jl,r}$ as the probability of a subject from subpopulation $s$ having a consumption level $r$ from food item $j$ given allocation to local dietary cluster $L_{ij}=l$. 

The cluster-specific parameters were each drawn from a flat, symmetric Dirichlet distribution, $\theta_{0jh,\cdot} = \{\theta_{0jh,1},\ldots,\theta_{0jh,d}\} \sim Dir_d(\eta)$, $\theta^{(s)}_{1jl,\cdot}=\{\theta^{(s)}_{1jl,1},\ldots,\theta^{(s)_{1jl,d}}\} \sim Dir_d(\eta)$, where $\eta=1$. The hyperparameters of the Beta-Bernoulli process component of the RPC were drawn from a gamma prior, $\beta^{(s)} \sim \Gamma(1, 1)$. To encourage a less informative beta-Bernoulli process, $\beta^{(s)}$ was set to 1 for simplicity. 


\subsection{MCMC Performance}
Following the simulation, inference is based on an MCMC run of 20,000 iterations, after a 5,000 burn-in. Given the tendency, acknowledged in the simulation study, of parameters gravitating to a preferred node and remaining there for subsequent iterations in large samples, the random permutation sampler was applied to encourage mixing \citep{fruhwirth2001markov}. Furthermore, the overfitted model is also prone to generating extraneous and redundant clustering \citep{van2015overfitting}. Redundancies were removed by creating a posterior pairwise comparison matrix based on the full MCMC output. Hierarchical clustering was then performed using the complete linkage approach, restricting to the median number of nonempty clusters larger than 5\% in size \citep{krebs1989ecological, medvedovic2002bayesian}. This threshold was determined from the simulation study in order to focus on identifying the clusters of global interest. The trace plots of $\beta^{(s)}$ and $\pi_{1:K}$ showed a good mixing and rapid convergence. All model parameters were estimated by calculating the posterior median and 95\% credible intervals. 

\subsection{NBDPS Results}
\subsubsection{RPC Results}
\begin{figure}[H]
\centering
\includegraphics[width=0.45\textwidth, height=0.95\textheight]{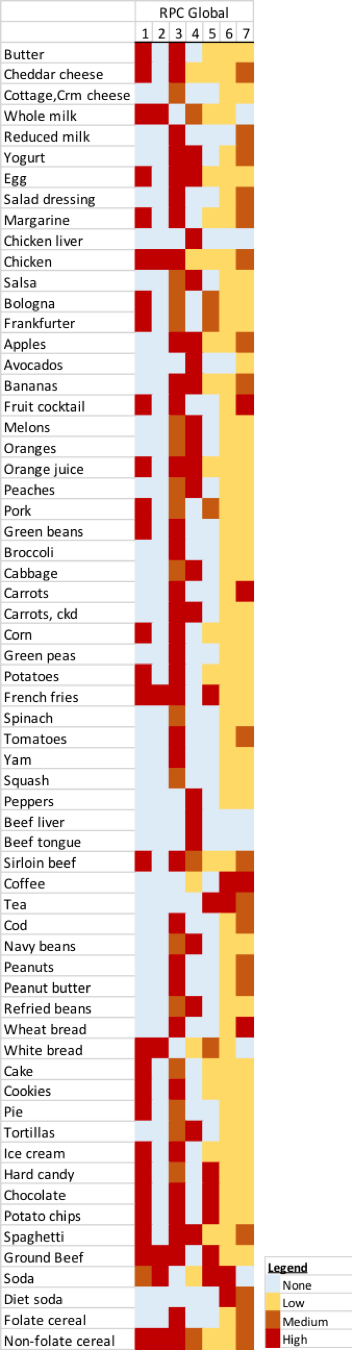}
\caption{Heat map showing modal consumption level of RPC global profiles. Legend: Blue Ð no consumption, Yellow Ð low consumption, Orange Ð medium consumption, Red Ð high consumption. \label{fig:profiles}}
\end{figure}

\begin{figure}[H]
  \centering
    \includegraphics[width=\textwidth, height=0.5\textheight]{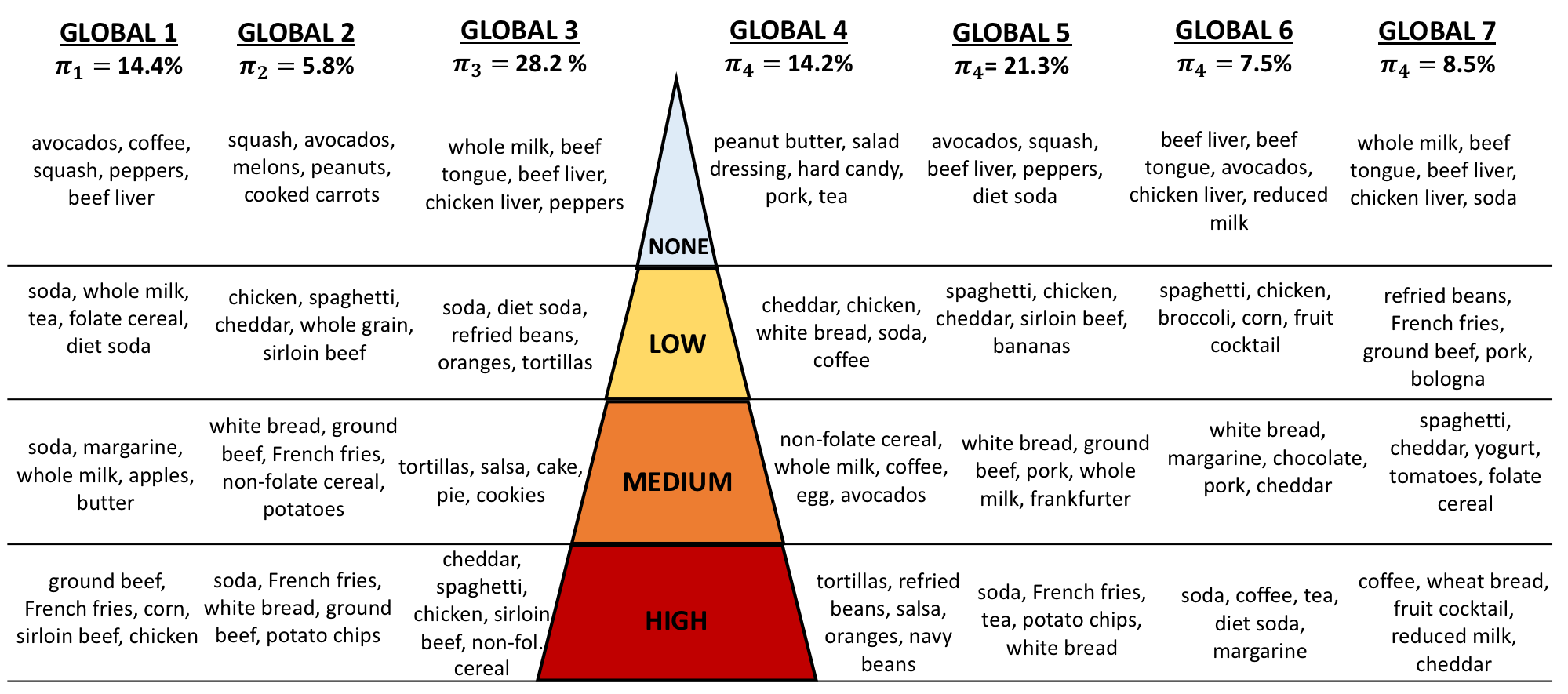}\\
      \caption{Foods most likely to be consumed at each consumption level of global profiles. Level of consumption decreases as you move from bottom to top. \label{fig:rpcpyramids}}
\end{figure}

The RPC model identified a total of seven nonempty global cluster patterns. A heat map illustrating the patterns of the global profiles is provided in Figure \ref{fig:profiles}.  Global profile behaviors were described by foods most likely to have a given consumption level within each cluster. Figure \ref{fig:rpcpyramids} illustrates the top five foods with the highest probability of having a given consumption level for each global profile. Foods are listed from top to bottom in order of frequency of consumption, with the largest level of consumption illustrated at the bottom. Global profile 1 had a high consumption of meats and fatty foods (candy, chocolate, beef, chicken starches). Global profile 2 had a high consumption of fast foods (soda, white bread, french fries, ground beef, potato chips). Global profile 3 had a high consumption of chicken, cheese, and beef.  Foods commonly found in a ``Tex-Mex'' or Latino style diet were likely to be consumed at a high level in global profile 4. Global profile 5 had a high consumption of ``snack-style'' foods (soda, french fries, tea, potato chips) as well as a medium consumption of pork and frankfurters.  Global profile 6 had a high consumption of caffeine products (soda, coffee, tea). Global profile 7 was the most prudent profile with highly consumed foods including wheat bread, fruit cocktail, and low-fat milk.


\begin{figure}[H]
  \centering
    \includegraphics[width=\textwidth, height=0.4\textheight]{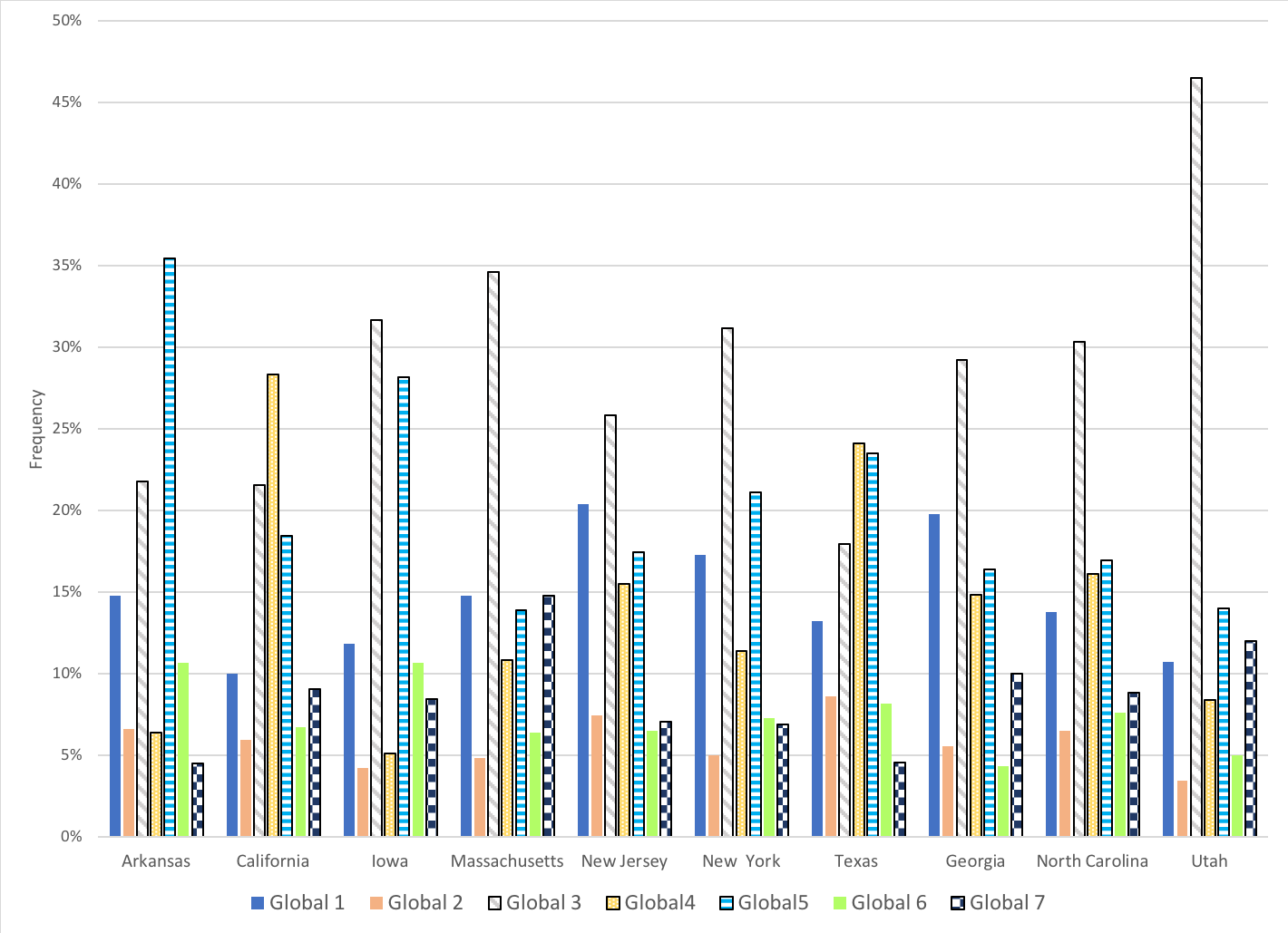}
      \caption{Frequency distribution of global profiles by subpopulation (state) for the global profiles identified in the model. Global profiles are displayed in numerical order from left to right, with corresponding colors indicated in the legend below bar plots \label{fig:globalrpc}}
\end{figure}

The distribution of these profiles by each subpopulation is illustrated in Figure \ref{fig:globalrpc}. 
Global profile 3 (gray) was the largest and most prominent profile in all states except for Arkansas and California. In Arkansas, global profile 5 was most prominent. Texas had a strong representation of global profile 4, which had a high consumption of foods commonly found in a `Tex-Mex' style diet (tortillas, refried beans, salsa).

\begin{figure}[H]
  \centering
    \includegraphics[width=0.5\textwidth, height=0.75\textheight]{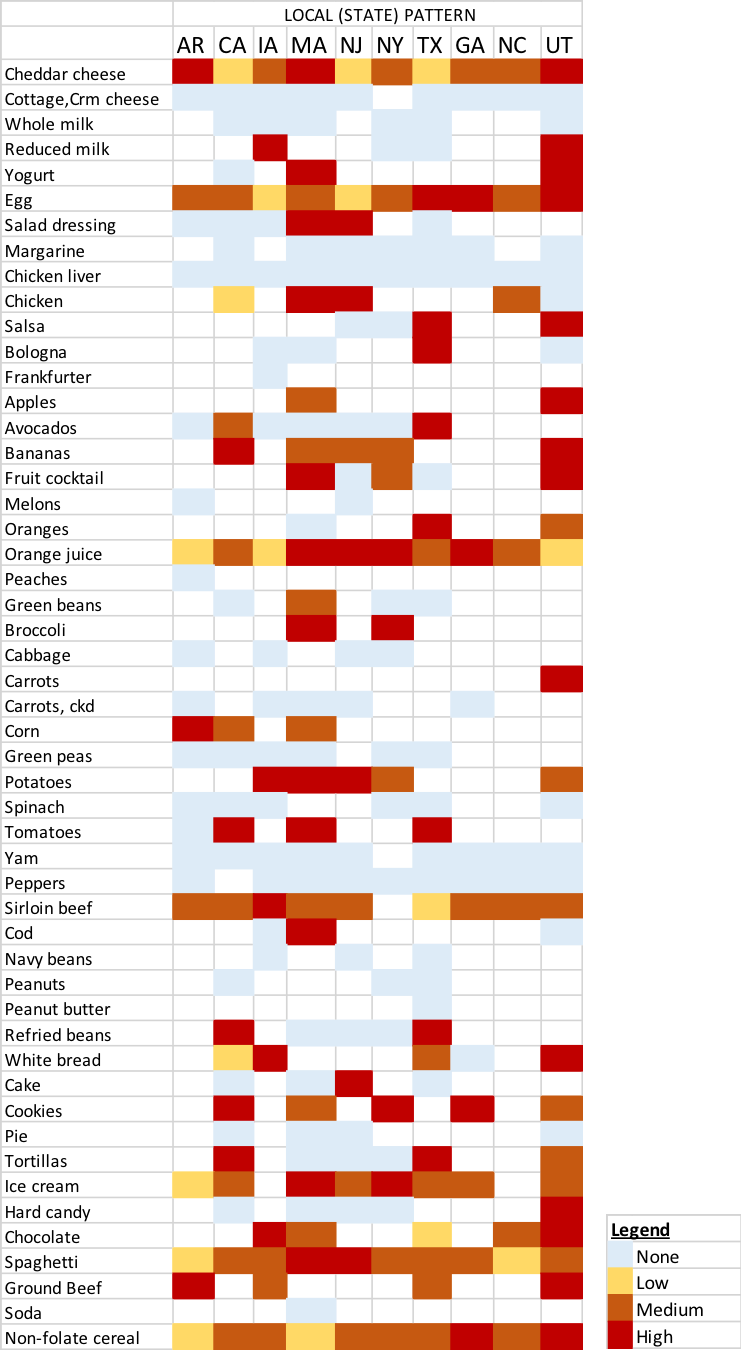}\\
      \caption{Heat map illustrating foods with a tendency to deviate from the global profile $(\nu_j^{(s)} < 0.5)$, by subpopulation. Legend: Blue Ð no consumption, Yellow Ð low consumption, Orange Ð medium consumption, Red Ð high consumption.\label{RPCgrid}}
\end{figure}

While the food patterns found at the global level were shared amongst subjects from different subpopulations, unique behaviors were more pronounced within each state at the local level. A subset of foods were found to deviate at the subpopulation level for all ten states. A food with a tendency to deviate in favor of a subpopulation pattern was identified as $\nu_j^{(s)} < 0.50$.  Eight foods were found to share a tendency to deviate in favor of non consumption in all ten states (butter, squash, beef liver, beef tongue, coffee, tea, diet soda, folate cereal). Many of these foods were found to have consumption levels at the global level. This implies that subjects either did consume according to their global profile pattern, or did not consume that food at all. Four foods had a tendency to not deviate to any of the local clusters, and remain in accordance with the global profiles (pork, french fries, potato chips, wheat bread). All other food variables with varying consumption patterns by state, are illustrated in Figure \ref{RPCgrid}. At the local level, foods aggregated to a single cluster $(\lambda^{(s)}>0.95)$, indicating less variability of consumption patterns within each subpopulation. 

States with higher representation of Latino populations (Texas and California) showed a tendency for higher consumption of foods commonly found in a Latino diet (tortillas, refried beans, salsa, avocados), whereas states such as Massachusetts, New Jersey, and New York showed a tendency toward lower consumption of those foods. Another highly discriminating food was spaghetti, which was consumed at a high level for those subjects in global profile 3. This was the most populous global profile in North Carolina and Utah, yet subjects from those states were likely to deviate in favor of a lower consumption level, respectively. Similarly, subjects from Massachusetts and New Jersey who were not allocated to that global profile (i.e. allocated instead to global profile 2 and global profile 5) still favored a high consumption of spaghetti. Utah subjects had a tendency to deviate in favor of a high consumption of child-friendly fruits (apples, bananas, fruit cocktail) and dairy (egg, reduced fat milk). 

\subsubsection{Comparisons with Other Models}\label{sec:nbdpsprev}

We also analyzed the NBDPS data using the previous methods. In general, these models had an exhaustive number of clusters that shared patterns reflected in both the global and local clusters of the RPC. Using the posterior median estimates, NBDPS participants were classified into the cluster with the highest posterior allocation probability. The mean and 95\% credible interval of these probabilities is summarized in Table \ref{tab:postp}. The LCA-4 model had the strongest allocation probabilities, and the DPM model had the weakest. 

\begin{table}
\centering
\begin{tabular}{ l c c}
\hline
& \bf{Non-empty Clusters} & {\textbf{Posterior Probability of Assigned Cluster}} \\
 & $(K_0)$ & Mean  (95\% Credible Interval)\\
  \hline
tLCA & 4 & .999  (.56, 1.00)\\ 
DPM & 16 & 0.934 (0.44, 1.00) \\ 
oFMM &  15 & 0.974  (0.49,1.00 )  \\
$LPP_2$ & 9 & 0.964 (0.44 , 1.00)  \\ 
RPC & 7 & 0.999 (0.52, 1.00)  \\ 

\end{tabular}
\caption{Summary comparing number of nonempty clusters and mean posterior probability of NBDPS assigned cluster  95\% credible interval of maximum posterior probability of subject allocation to preferred cluster \label{tab:postp}}
\end{table}

\begin{figure}[H]
  \centering
    \includegraphics[width=0.8\textwidth, height=0.9\textheight]{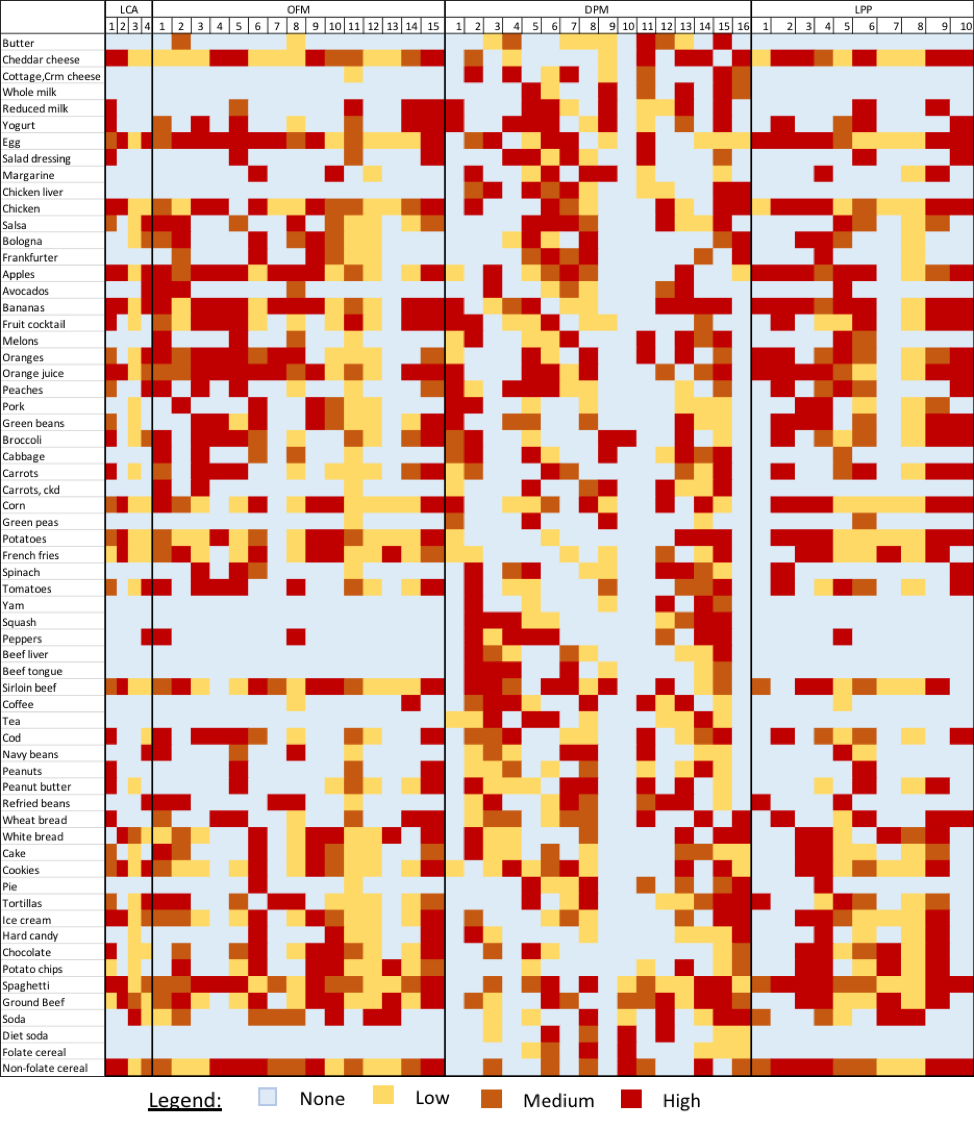}\\
     \caption{Heat map showing modal consumption level of each of the nonempty clusters derived from the LCA-4, OFM-15, DPM-16, $LPP_2$-10. Blue - no consumption, yellow - low consumption, orange - medium consumption, red - high consumption.\label{fig:prevmeth}}
\end{figure}

Despite strong allocation probabilities, overgeneralization was a major limitation of the LCA-4 model. Over 20 foods were identified as favoring no consumption across all four classes. For example, pie favored no consumption for all cluster patterns (Appendix C, Figure \ref{fig:prevmeth}). Yet, the RPC model showed that this tendency for non-consumption was only prominent in California, Massachusetts, New Jersey, and Utah (Figure \ref{RPCgrid}). The remaining subpopulations favored corresponding  global profile patterns which indicated a non-consumption in only three of the seven profiles. Both the OFM and DPM models derived an excessive number of clusters ($K_{OFM} = 15, K_{DPM}=16$). In the DPM model, only one derived cluster was larger than 5\% in size. Like the LCA-4 model, the OFM modal cluster pattern indicated a general non-consumption of nine foods that was not reflected in the RPC model. For example, the consumption of yams by the state of New York is masked in the OFM model. This consumption is relevant to New York participants that belong to global profiles 3, 6, or 7 that favor a high or low consumption of yams. The $LPP_2$ was the only model that allows the global clustering assumption to be relaxed with global and local clustering. A total of nine nonempty clusters were derived in the $LPP_2$, but none of the foods in the dataset favored a deviation away from the global profile patterns $(\nu_j > 0.99: \text{for all }j \in 1,\ldots, p)$. By allowing deviations by the known subpopulation, the RPC was able to clearly identify and distinguish deviating behaviors from the global profile trends. 

\section{Discussion} 

The RPC method provides a convenient and informative population-based model that is able to adapt and account for potential deviations occurring within subpopulations. It creates a practical solution to the global clustering problem by allowing certain variables to assume a distribution separate from their assigned overall cluster. This separate distribution is unique and can be shared among participants belonging to the same subpopulation regardless of the overall cluster assigned to those participants. A logistic regression model could be placed on relevant probabilities as desired to include covariates. 


The overfitted mixture framework of the RPC is sensitive to the selection of Dirichlet hyperpriors. Smaller Dirichlet hyperprior values can slow down the rate of cluster growth \citep{rousseau2011asymptotic}. Our selection of $\frac{1}{K}$ as the hyperprior provided a modest rate of growth for our large sample population size. Other more informative or even nonsymmetric hyperpriors can be used as an alternative, as deemed appropriate. Additionally, the overfitted finite mixture model is still prone to redundancies in the cluster profile patterns. This often can require an additional step of post-processing to examine the final profile results and remove redundant patterns as they are expressed in the sampler. An alternative would be to use a hierarchical hybrid Dirichlet that can handle discrete data, if a more complex approach is desired. This extension is a different potential direction that to our knowledge has not yet been pursued.

Nevertheless, RPC allows researchers to characterize behaviors attributable to a general population or isolated within a subpopulation. In the application using NBDPS dietary data, numerous regional food trends were identified using the RPC model that were not apparent using other approaches.
\newpage
\section*{Appendix A: Simulation Study Details}
Global and local cluster patterns were generated by applying a probability of 0.7 to the expected true response level and a probability of 0.1 to all other response categories. Let $r^*$ denote the expected response level for a given global cluster. If $r^* = 3$ for variable $j=1$, then $(\theta_{011,1},\ldots,\theta_{011,d}) = (0.1, 0.1, 0.7, 0.1)$. A table of expected location of $r^*$ mode for each global cluster profile is provided below.

\subsection*{Case 1: Strictly Global}
Each subject $i$ belongs to one of $S= 4$ subpopulations$(s_i \in 1,2,3,4)$. Within that subpopulation, each subject is assigned to one of 3 global profiles $(C_i \in 1,2,3)$. Probability of variable deviation is preset to $\nu_j^{(s)} = 1$, such that $G_{ij}=1$ for all subjects $i \in (1,\ldots, n)$, and all variables $j \in (1,\ldots, p)$. Expected global profile pattern, defined by modal response, is provided in Table \ref{tab:glob1sim}. \\

Subject-level observed data was generated as follows: \begin{algorithm}
\caption{Simulation Case 1} \label{sim1alg}
  \begin{algorithmic}[1]
        \STATE Generate $y_{ij}| C_i, s_i \sim \text{Multinomial}(\theta_{0jC_i,1},\ldots,\theta_{0jC_i,d})$
  \end{algorithmic}
\end{algorithm}

\begin{table}[H]
{\small \renewcommand{\arraystretch}{.8}
\begin{tabular}{|c|c|c|c||c|c|c|c|}
\toprule
$j$&Global&Global&Global & $j$& Global&Global&Global\\
 & 1 & 2& 3 & & 1 &2 &3 \\
\hline
1 & 3 & 2 &1 & 26 & 1 & 4& 2 \\
2 & 3 & 2 &1 & 27 & 1 & 4 & 2 \\
3 & 3 & 2 &1 & 28 & 1 & 4 & 2 \\
4 & 3 & 2 &1 & 29 & 1 & 4 & 2 \\
5 & 3 & 2 &1 & 30 & 1 & 4 & 2 \\
6 & 3 & 2 &1 & 31 & 1 & 4 & 3 \\
7 & 3 & 2 &1 & 32 & 1 & 4 & 3 \\
8 & 3 & 2 &1 & 33 & 1 & 4 & 3 \\
9 & 3 & 2 &1 & 34 & 1 & 4 & 3 \\
10 & 3 & 2 &1 & 35 & 1 & 4 & 3 \\
11 & 3 & 4 & 2 & 36 & 1 & 4 & 3 \\
12 & 3 & 4& 2 & 37 & 1 & 4 & 3 \\
13 & 3 & 4& 2 & 38 & 1 & 4 & 3 \\
14 & 3 & 4& 2 & 39 & 1 & 4 & 3 \\
15 & 3 & 4& 2 & 40 & 1 & 4 & 3 \\
16 & 3 & 4& 2 & 41 & 1 & 4 & 3 \\
17 & 3 & 4& 2 & 42 & 1 & 4 & 3 \\
18 & 3 & 4& 2 & 43 & 1 & 4 & 3 \\
19& 3 & 4& 2 & 44 & 1 & 4 & 3 \\
20& 3 & 4& 2 & 45 & 1 & 4 & 3 \\
21 & 3 & 4& 2 & 46 & 1 & 4 & 3 \\
22& 3 & 4& 2 & 47 & 1 & 4 & 3 \\
23& 3 & 4& 2 & 48 & 1 & 4 & 3 \\
24& 3 & 4& 2 & 49 & 1 & 4 & 3 \\
25& 3 & 4& 2 & 50 & 1 & 4 & 3 \\
\bottomrule

\end{tabular}
}
\caption{Location of mode for each respective global cluster where maximal probability ($\theta_{0jC_i,r^*} = 0.7)$. \label{tab:glob1sim}}
\end{table}

\subsection*{Case 2: Strictly Local}
Each subject $i$ belongs to one of $S=8$ subpopulations $(s_i \in 1,2,\ldots, 8)$. Each subpopulation contains one local cluster. Probability of variable deviation is preset to $\nu_j^{(s)} = 0$, such that $G_{ij}=0$ for all subjects $i \in (1,\ldots, n)$, and all variables $j \in (1,\ldots, p)$. Expected local profile pattern, defined by modal response, is provided in Table \ref{tab:local1sim}.\\

Subject-level observed data was generated as follows:  
\begin{algorithm}
\caption{Simulation Case 2}\label{sim2alg}
  \begin{algorithmic}[2]
        \STATE Generate $y_{ij}| L_{ij}, s_i \sim \text{Multinomial}(\theta^{(s_i)}_{1jL_{ij},1},\ldots,\theta^{(s_i)}_{1jL_{ij},d})$
  \end{algorithmic}
\end{algorithm}

\begin{table}[H]
{\tiny\renewcommand{\arraystretch}{.8}
\resizebox{!}{.3\paperheight}{%
\begin{tabular}{|c|c|c|c|c|c|c|c|c|}
\toprule
$j$&Local&Local&Local&Local&Local&Local&Local&Local\\
  & 1 &2&3&4&5&6&7&8 \\
\hline
1 & 1 &1 & 2 & 2 & 3 & 3 &4 &4 \\ \hline
2 & 1 &2 &2 &4 &3 & 1 &4 &3 \\ \hline
3 & 1 & 3 & 2 & 1 &3 & 4 &4 & 2 \\ \hline
4 & 1 &4 &2 &3 &3 &2 &4 &1  \\ \hline
5 & 1 &1&2 &2& 3 &3& 4&4\\ \hline
6 & 1 &2 &2&4&3&1&4&3 \\ \hline
7 &1 &3&2&1&3&4&4&2 \\ \hline
8 & 1&4&2&3&3&2&4&1\\ \hline
9 &1&1&2&2&3&3&4&4\\ \hline
10 & 1 &2&2&4&3&1&4&3\\ \hline
11 & 1& 3& 2&1&3&4&4&2\\ \hline
12&1&4&2&3&3&2&4&1\\ \hline
13&1&1&2&2&3&3&4&4\\ \hline
14&1&2&2&4&3&1&4&3\\ \hline
15&1&3&2&1&3&4&4&2\\ \hline
16&1&4&2&3&3&2&4&1\\ \hline
17&1&1&2&2&3&3&4&4\\ \hline
18&1&2&2&4&3&1&4&3\\ \hline
19&1&3&2&1&3&4&4&2\\ \hline
20&1&4&2&3&3&2&4&1\\ \hline
21&1&1&2&2&3&3&4&4\\ \hline
22&1&2&2&4&3&1&4&3\\ \hline
23&1&3&2&1&3&4&4&2\\ \hline
24&1&4&2&3&3&2&4&1\\ \hline
25&1&1&2&2&3&3&4&4\\ \hline
26&1&2&2&4&3&1&4&3\\ \hline
27&1&3&2&1&3&4&4&2\\ \hline
28&1&4&2&3&3&2&4&1\\ \hline
29&1&1&2&2&3&3&4&4\\ \hline
30&1&2&2&4&3&1&4&3\\ \hline
31&1&3&2&1&3&4&4&2\\ \hline
32&1&4&2&3&3&2&4&1\\ \hline
33&1&1&2&2&3&&4&4\\ \hline
34&1&2&2&4&3&1&4&3\\ \hline
35&1&3&2&1&3&4&4&2\\ \hline
36&1&4&2&3&3&2&4&1\\ \hline
37&1&1&2&2&3&3&4&4\\ \hline
38&1&2&2&4&3&1&4&3\\ \hline
39&1&3&2&1&3&4&4&2\\ \hline
40&1&4&2&3&3&2&4&1\\ \hline
41&1&1&2&2&3&3&4&4\\ \hline
42&1&2&2&4&3&1&4&3\\ \hline
43&1&3&2&1&3&4&4&2\\ \hline
44&1&4&2&3&3&2&4&1\\ \hline
45&1&1&2&2&3&3&4&4\\ \hline
46&1&2&2&4&3&1&4&3\\ \hline
47&1&3&2&1&3&4&4&2\\ \hline
48&1&4&2&3&3&2&4&1\\ \hline
49&1&1&2&2&3&3&4&4\\ \hline
50&1&1&2&2&3&3&4&4\\ 
\bottomrule
\end{tabular}
}}
\caption{Location of maximal probability for each variable in respective local cluster where maximal probability ($\theta^{(s)}_{1jL_{ij},r^*} = 0.7)$.\label{tab:local1sim}}
\end{table}

\subsection*{Case 3: Hybrid}
Each subject $i$ belongs to one of $S=4$ subpopulations $(s_i \in 1,2,3,4)$. Within each subpopulation, each subject is assigned to one of two local clusters, $L_{ij} = l \in (1,2)$ and one of three global clusters, $C_i = k \in (1,2,3)$. Expected global profile patterns, defined by modal response, are provided in Table \ref{tab:glob1sim}, when $G_{ij}=1$. Expected local profile patterns, defined by modal response, are provided in Table \ref{tab:local1sim}, when $G_{ij}=0$. Probability of allocation to global or local was fixed for all variables within a subpopulation: \\
Subpopulation 1: $\nu_j^{(1)} = 0.25$ for all $j \in (1,\ldots,p)$, local profiles 1 and 2 (Table \ref{tab:local1sim})\\
Subpopulation 2: $\nu_j^{(2)} = 0.50$ for all $j \in (1,\ldots,p)$, local profiles 3 and 4 (Table \ref{tab:local1sim} \\
Subpopulation 3: $\nu_j^{(3)} = 0.75$ for all $j \in (1,\ldots,p)$, local profiles 5 and 6 (Table \ref{tab:local1sim} \\
Subpopulation 4: $\nu_j^{(4)} = 1.00$ for all $j \in (1,\ldots,p)$, local profiles 7 and 8 (Table \ref{tab:local1sim} \\

\begin{algorithm}
\caption{Simulation Case 3}\label{sim3alg}
  \begin{algorithmic}[3]
        \STATE Generate $G_{ij}|s_i \sim \text{Bernoulli}(\nu_j^{(s_i)})$
        		\IF{ $G_{ij}=1$ }
                		\STATE Generate $y_{ij} | C_i \sim \text{Multinomial}(\theta_{0jC_i,1},\ldots,\theta_{0jC_i,d})$
            	\ELSIF{$G_{ij}=0$}
                		\STATE Generate $y_{ij} | L_{ij}, s_i \sim \text{Multinomial}(\theta^{(s_i)}_{1jL_{ij},1},\ldots,\theta^{(s_i)}_{1jL_{ij},d})$
            	\ENDIF
  \end{algorithmic}
\end{algorithm}

\subsection*{Case 4: null set }
Each subject $i$ was assigned to one of $S=4$ subpopulations $(s_i \in 1,2,3,4)$. Subject-level observed data was randomly assigned from a discrete uniform distribution containing $d=4$ possible response levels. 
\begin{algorithm}
\caption{Simulation Case 4}
  \begin{algorithmic}[4]
        \STATE Generate $y_{ij} \sim \text{Uniform}(1,\ldots,d)$
  \end{algorithmic}
\end{algorithm}

\subsection*{Case 5: Mock NBDPS}
Each subject $i$ was assigned to one of $S=10$ subpopulations $(s_i \in 1,2,\ldots, 10)$. Within each subpopulation, each subject is assigned to a global profile, labelled $C_i$, and each subpopulation contained at least one of 28 unique profiles containing a combination of global and local variables. Subject-level observed data was generated as referenced. Subpopulations 1-3, and 8 contained 1200 subjects, with 400 subjects equally assigned to one of three global profiles. Each of these subpopulations contained 2 local profiles with subjects evenly assigned to each profile (600 subjects per local profile for subpopulations 1-3; 400 subjects per local profile for subpopulation 8). Subpopulations 4-6, and 9 contained 1200 subjects with 400 subjects equally assigned to one of the three global profiles (400 subjects per global profile). Subpopulation 10 contained 1600 subjects, where 400 subjects contained one global profile and no local profile and 1200 subjects with one local profile and no global profile. Expected global profile patterns, defined by modal response, are provided in Table \ref{tab:glob1sim}, when $G_{ij}=0$.  Expected local profile patterns, defined by modal response, are provided in Table \ref{tab:case5nu}, when $G_{ij}=1$.  

\begin{table}
{\tiny\renewcommand{\arraystretch}{.8}
\resizebox{!}{.3\paperheight}{%
\begin{tabular}{|c|c|c|c|c|c|c|c|c|c|c|c|c|c|}
\toprule
$j$&\multicolumn{2}{|c|}{Subpop 1}&\multicolumn{2}{|c|}{Subpop 2}&\multicolumn{2}{|c|}{Subpop 3}&\multicolumn{2}{|c|}{Subpop 7}&\multicolumn{2}{|c|}{Subpop 8}&\multicolumn{2}{|c|}{Subpop 9}&Subpop 10\\
  & Local 1 &Local 2& Local 1 &Local 2&Local 1 &Local 2&Local 1 &Local 2&Local 1 &Local 2&Local 1 &Local 2&Local 1 \\
\hline
1 & 2 &2 & 3 & 2 &  &  &4 &2 & & & 2&2&4 \\
\hline
2 & 1 &4&1&4&2&2&1&4&&&3&1&4\\
\hline
3 & & & & & 4&3& 1&3&1&2&&&3\\
\hline
4 & 3&3&3&1&&&2&1&&&1&1&3  \\
\hline
5 & &&&&1&3&4&2&&&4&2&4\\
\hline
6 & & &2&2&3&2&3&4&&&2&1&2 \\
\hline
7 &&&&&&&3&3&&&3&1&3 \\
\hline
8 & &&3&2&2&1&3&4&&&2&1&2\\
\hline
9 &&&3&3&1&1&1&1&2&3&1&4&2\\
\hline
10 & &&4&2&4&4&1&1&&&3&4&2\\
\hline
11 & &&2&2&&&4&4&&&2&3&3\\
\hline
12&&&4&4&4&3&4&3&&&1&4&3\\
\hline
13&&&2&1&4&1&4&2&&&2&1&4\\
\hline
14&&&3&2&3&3&1&1&2&4&2&3&4\\
\hline
15&&&3&1&4&3&4&3&&&3&2&3\\
\hline
16&&&1&1&1&3&2&3&&&2&2&2\\
\hline
17&&&1&1&2&2&3&4&&&1&3&3\\
\hline
18&&&4&3&4&1&1&1&&&1&4&2\\
\hline
19&2&4&4&3&&&4&4&&&4&1&4\\
\hline
20&&&4&1&1&1&3&3&&&&&2\\
\hline
21&&&&&1&1&1&1&&&3&1&2\\
\hline
22&3&4&&&4&4&4&3&&&3&3&2\\
\hline
23&1&4&2&4&1&3&3&1&&&3&3&2\\
\hline
24&&&3&1&2&3&4&2&&&4&4&3\\
\hline
25&2&2&&&1&2&3&3&&&4&4&3\\
\hline
26&&&3&4&2&1&3&1&&&3&3&4\\
\hline
27&&&&&1&4&2&4&&&3&3&2\\
\hline
28&&&&&3&3&2&3&&&4&1&3\\
\hline
29&&&3&4&3&4&4&1&&&2&2&3\\
\hline
30&&&4&3&4&1&1&1&&&2&4&2\\
\hline
31&&&&&4&4&4&1&2&1&&&4\\
\hline
32&&&&&4&1&1&2&2&2&2&3&1\\
\hline
33&&&2&2&3&2&2&1&&&4&2&4\\
\hline
34&&&&&4&1&4&1&&&1&1&1\\
\hline
35&4&4&&&2&1&1&1&4&4&2&4&4\\
\hline
36&&&1&2&3&2&2&1&&&1&2&4\\
\hline
37&&&1&1&1&2&1&2&&&3&4&2\\
\hline
38&&&4&4&&&2&2&&&4&2&2\\
\hline
39&&&2&1&1&4&4&2&&&1&2&2\\
\hline
40&&&&&2&1&2&3&&&4&2&4\\
\hline
41&&&2&4&2&4&3&3&&&3&1&1\\
\hline
42&&&&&3&2&3&1&&&4&1&3\\
\hline
43&1&2&4&2&3&1&4&2&&&3&1&3\\
\hline
44&&&&&1&4&3&3&&&3&1&1\\
\hline
45&&&3&4&&&4&4&&&&&4\\
\hline
46&&&2&2&3&4&3&1&&&1&2&2\\
\hline
47&&&&&1&3&3&4&&&3&2&2\\
\hline
48&&&&&4&1&3&4&&&2&4&2\\
\hline
49&1&3&&&1&1&3&2&&&3&1&3\\
\hline
50&2&3&1&4&4&4&1&3&&&&&4\\
\bottomrule
\end{tabular}
}}
\caption{Case5: Location of maximal probability for each variable in respective local cluster, where $(\theta^{(s)}_{1jL_{ij},r^*}=0.7)$ \label{tab:case5nu}}
\end{table}

\begin{algorithm}[H]
\caption{Simulation Case 5}
  \begin{algorithmic}[5]
        \STATE Subpopulation 1: Generate $y_{ij}$ as in Algorithm \ref{sim3alg}, where $\nu_j^{(s)} = 0.75$.
        \STATE Subpopulation 2: Generate $y_{ij}$ as in Algorithm \ref{sim3alg}, where $\nu_j^{(s)} = 0.50$.
         \STATE Subpopulation 3: Generate $y_{ij}$ as in Algorithm \ref{sim3alg}, where $\nu_j^{(s)} = 0.25$.
         \STATE Subpopulation 4-6: Generate $y_{ij}$ as in Algorithm \ref{sim1alg}.
  	 \STATE Subpopulation 7: Generate $y_{ij}$ as in Algorithm \ref{sim2alg}.
	 \STATE Subpopulation 8: Generate $y_{ij}|C_i = 1$ as in Algorithm \ref{sim2alg} and $y_{ij}|C_i = 2,3$ as in algorithm \ref{sim3alg}, where $\nu_j^{(s)} = 0.9$.
	 \STATE Subpopulation 9: Generate $y_{ij}$ as in Algorithm \ref{sim1=3alg}, where $\nu_j^{(s)} = 0.1$.
	 \STATE Subpopulation 10: Generate $y_{ij}|C_i =1$ as in Algorithm \ref{sim1alg} and $y_{ij}|C_i =2$ as in Algorithm \ref{sim2alg}.
  \end{algorithmic}
\end{algorithm}

\subsection*{Case 6: Violations of Beta-Bernoulli assumption}
Each subject $i$ belongs to one of $S=4$ subpopulations $(s_i \in 1,2,3,4)$. Within each subpopulation, each subject is assigned to one of two local clusters, $L_{ij} = l \in (1,2)$ and one of three global clusters, $C_i = k \in (1,2,3)$. Expected global profile patterns, defined by modal response, are provided in Table \ref{tab:glob1sim}, when $G_{ij}=1$. Expected local profile patterns, defined by modal response, are designed as in Case 3, with the 8 local profile patterns distributed across the four subpopulations. Probability of deviation to global or local level for all variables within a subpopulation was drawn from a different subpopulation: \\

\begin{algorithm}
\caption{Simulation Case 6}
  \begin{algorithmic}[6]
      		\IF{$s_i = 1$} 
	\STATE Randomly draw $\nu_j^{(1)} \sim \text{Beta}(2,1)$	 for all $j \in (1,\ldots, p)$
		
		\ELSIF{$s_i = 2$}
	\STATE Randomly draw $\nu_j^{(2)} \sim U(0,1)$ for all $j \in (1, \ldots, p)$
		
		\ELSIF{$s_i = 3$}
	\STATE Randomly draw $\nu_j^{(3)} \sim P(N(0,1))$ for all $j \in (1,\ldots, p)$
		\ELSIF{$s_i = 4$}
	\STATE Randomly draw $\nu_j^{(3)}$ from a standard Cauchy distribution for all $j \in (1,\ldots, p)$
		\ENDIF
		
        		\IF{ $G_{ij}=1$ }
                		\STATE Generate $y_{ij} | C_i \sim \text{Multinomial}(\theta_{0jC_i,1},\ldots,\theta_{0jC_i,d})$
            	\ELSIF{$G_{ij}=0$}
                		\STATE Generate $y_{ij} | L_{ij}, s_i \sim \text{Multinomial}(\theta^{(s_i)}_{1jL_{ij},1},\ldots,\theta^{(s_i)}_{1jL_{ij},d})$
            	\ENDIF
  \end{algorithmic}
\end{algorithm}

\subsection*{Case 7: Continuous Data}
Each subject $i$ belongs to one of $S=2$ subpopulations $(s_i \in 1,2)$ of size 1500 subjects each, describing $p=30$ variables. Within each subpopulation, each subject is assigned to one of two local clusters, $L_ij = l \in (1,2)$ and one of two global clusters, $C_i = k \in (1,2)$. Allocation to global or local level for all variables within a subpopulation was preset such that following variables were allocated from the local density from subpopulation 1: vars (6, 7, 16, 17, 26, 27);  subpopulation 2: vars (5, 8, 15, 18, 25, 28). Subject-level observed data for  global and local profiles are drawn from a normal distribution with respective means, $\mu_0, \mu^{(s)}_1$. Standard deviations remained uniform depending on the case: (a) $\sigma = 0.1$, (b) $\sigma = 1.0$ (c) $\sigma = 3.0$. 

\begin{table}[H]
\begin{tabular}{|l|c|c|}

\hline
Case & Global Density & Local Density\\
 & $N(\mu_0,\sigma)$ & $N(\mu_1^{(s)}, \sigma)$ \\
\hline
7a: Completely Separate & $N(-9,0.1)$ & 1: $N(5,0.1) \quad N(9, 0.1) $ \\
	& $N(2,0.1)$ & 2: $N(-5,0.1) \quad N(-2,0.1) $\\
	\hline
7b: Partially Separate &  $N(-9,1)$ & 1: $N(5,1) \quad N(9, 1) $ \\
	& 2: $N(2,1)$ & $N(-5,1) \quad N(-2,1) $\\
\hline
7c: Complete Overlap &  $N(-9,3)$ & 1: $N(5,3) \quad N(9, 3) $ \\
	& 2: $N(2,3)$ & $N(-5,3) \quad N(-2,3) $\\

	\hline
\end{tabular}
\caption{Simulation set-up for each sub-case for continuous set from a $N(\mu,\sigma)$ distribution. Each dataset contains 2 subpopulations with $p=30$ variables. Each subpopulation contains information from 2 global and 2 local clusters. \label{tab:sim7}}
\end{table}

\begin{algorithm}
\caption{Simulation Case 7}
  \begin{algorithmic}[7]
        \STATE Generate $G_{ij}|s_i \sim \text{Bernoulli}(\nu_j^{(s_i)})$
        		\IF{ $G_{ij}=1$ }
                		\STATE Generate $y_{ij} | C_i \sim N(\mu_0,\sigma)$
            	\ELSIF{$G_{ij}=0$}
                		\STATE Generate $y_{ij} | L_{ij}, s_i \sim N(\mu^{(s)}_1,\sigma)$            	
		\ENDIF
  \end{algorithmic}
\end{algorithm}

\clearpage
\newpage
\section*{Appendix B: Supporting Figures from Mock NBDPS Simulation Study (Section \ref{sec:simresults})}\label{AppA}
Using one of the replicates from the mock NBDPS simulations (Section \ref{sec:mocksim}), the figure below provides a comparative illustration of variables that were identified as deviating from the global cluster pattern by each subpopulation. The figure on the left shows the variables that deviated as a result of the RPC model and the figure on the right illustrates the true deviated variables. Discrepancies in the predicted and actual results were found in subpopulation 7. This subpopulation contained patterns where all variables deviated from the global pattern and contained two unique local clusters. The RPC model misspecified the completely local pattern as a global pattern for subpopulation 7. Most of the deviation was identified as deviating to local for subpopulation 10. However, some locally allocated variables overlapped with the globally expected value and were therefore considered global in those cases. 

\newpage

\section*{Appendix C: Supporting Figure for NBDPS Previous Methods (Section \ref{sec:nbdpsprev})}\label{AppC}

\clearpage
\bigskip
\begin{center}
{\large\bf SUPPLEMENTARY MATERIAL}
\end{center}

An example dataset to illustrate the RPC method all source code, data, and expected output are publicly available at \url{https://github.com/bjks10/RPC}.

\begin{description}

\item[RPC source code:] MATLAB code to perform the RPC method and supporting files  can be found at \url{https://github.com/bjks10/RPC/SourceCode}.
\item[Simulated data set:] Simulated example data set to illustrate RPC method is available at \url{https://github.com/bjks10/RPC/ExampleData.mat}.
\item[Expected Output:] Execution of the RPC example should produce the following output found at \url{https://github.com/bjks10/RPC/Output}
\item[Posterior Computation:] MCMC Gibbs sampler for posterior computation of RPC 

\begin{enumerate}
\item Update the global component indicators $(G_{ij} \mid s_i=s) \sim \text{Bern}(p_{ij})$, where $$p_{ij}=\frac{\nu_j^{(s)}\prod_{r=1}^d \Theta_{0jC_i,r}^{\mathbf{1}(y_{ij}=r)}}{\nu_j^{(s)}\prod_{r=1}^d\Theta_{0jC_i,r}^{\mathbf{1}(y_{ij}=r)}+(1-\nu_j^{(s)})\prod_{r=1}^d (\Theta^{(s)}_{1jL_{ij},r})^{\mathbf{1}(y_{ij}=r)}}$$ for each subject $i \in (1,\ldots,n)$ with respective subpopulation index $s$. 

\item Update global cluster index $C_i$, $i=1,\ldots,n$ from its multinomial distribution where $$Pr(C_i=h) = \frac{\pi_h \prod_{j:G_{ij}=1} \prod_{r=1}^d \Theta_{0jh,r}^{\mathbf{1}(y_{ij}=r)}}{\sum_{l=1}^K \pi_l \prod_{j:G_{ij}=1} \prod_{r=1}^d \Theta_{0jl,r}^{\mathbf{1}(y_{ij}=r)}}.$$

\item Update local cluster index $L_{ij}$ for all $i:s_i=s$ and $j=1,\ldots,p$, repeating for each $s$, from its multinomial distribution conditional on $s_i=s$ where $$Pr(L_{ij}=h)=\frac{\lambda^{(s)}_h \prod_{r=1}^d \left(\Theta^{(s)}_{1jh,r}\right)^{\mathbf{1}(y_{ij}=r,G_{ij}=0)}}{\sum_{l=1}^K \lambda^{(s)}_l \prod_{r=1}^d \left(\Theta^{(s)}_{1jl,r}\right)^{\mathbf{1}(y_{ij}=r, G_{ij}=0)}}.$$

\item Update the global clustering weights $$\pi=(\pi_1,\ldots,\pi_K) \sim \text{Dir}\left(\frac{1}{K}+\sum_{i=1}^n \mathbf{1}(C_i=1),\ldots, \frac{1}{K}+\sum_{i=1}^n \mathbf{1}(C_i=K)\right).$$

\item Update the local clustering weights in subpopulation $s$, $$\lambda^{(s)}=\left(\lambda_1^{(s)},\ldots,\lambda_K^{(s)}\right) \sim \text{Dir}\left(\frac{1}{K} + \sum_{i:s_i=s} \sum_{j=1}^p \mathbf{1}(L_{ij}=1), \ldots, \frac{1}{K} + \sum_{i:s_i=s} \sum_{j=1}^p \mathbf{1}(L_{ij}=K)\right).$$

\item Update the multinomial parameters, where $\eta$ is a flat, symmetric Dirichlet hyperparameter preset at 1 
\begin{align*}
\theta_{0jh,\cdot} &\sim \text{Dir}\left(\eta+\sum_{i:G_{ij}=1,C_i=h} \mathbf{1}(y_{ij}=1),\ldots,\eta+\sum_{i:G_{ij}=1,C_i=h} \mathbf{1}(y_{ij}=d)\right) \\
\theta^{(s)}_{1jh,\cdot} &\sim \text{Dir}\left(\eta+\sum_{i:G_{ij}=0,L_{ij}=h,s_i=s} \mathbf{1}(y_{ij}=1),\ldots,\eta+\sum_{i:G_{ij}=0,L_{ij}=h,s_i=s} \mathbf{1}(y_{ij}=d)\right)
\end{align*}

\item Update $\nu_j^{(s)} \sim \text{Be}(1+\sum_{i:s_i=s} G_{ij}, \beta^{(s)} + \sum_{i:s_i=s} (1-G_{ij}))$.

\item Update Beta-Bernoulli hyperparameter: $\beta^{(s)} \sim \text{Ga}(a+p, b - \sum_{j=1}^p \log(1-\nu_j^{(s)}))$. 

\end{enumerate}

\end{description}


\bibliographystyle{apalike}
\bibliography{RPCRefs}

\begin{thebibliography}{}

\bibitem[Dunson, 2009]{dunson2009lpp}
Dunson, D.~B. (2009).
\newblock Nonparametric bayes local partition models for random effects.
\newblock {\em Biometrika}, 96(2):249--262.

\bibitem[Dunson and Xing, 2009]{dunson2009nonparametric}
Dunson, D.~B. and Xing, C. (2009).
\newblock Nonparametric bayes modeling of multivariate categorical data.
\newblock {\em Journal of the American Statistical Association},
  104(487):1042--1051.

\bibitem[Figueiredo and Jain, 2002]{figueiredo2002unsupervised}
Figueiredo, M. A.~T. and Jain, A.~K. (2002).
\newblock Unsupervised learning of finite mixture models.
\newblock {\em IEEE Transactions on Pattern Analysis and Machine Intelligence},
  24(3):381--396.

\bibitem[Ford et~al., 2010]{ford2010poly}
Ford, J.~D., Elhai, J.~D., Connor, D.~F., and Frueh, B.~C. (2010).
\newblock Poly-victimization and risk of posttraumatic, depressive, and
  substance use disorders and involvement in delinquency in a national sample
  of adolescents.
\newblock {\em Journal of Adolescent Health}, 46(6):545--552.

\bibitem[Fr{\"u}hwirth-Schnatter, 2001]{fruhwirth2001markov}
Fr{\"u}hwirth-Schnatter, S. (2001).
\newblock Markov chain monte carlo estimation of classical and dynamic
  switching and mixture models.
\newblock {\em Journal of the American Statistical Association},
  96(453):194--209.

\bibitem[Haraldsdottir, 1993]{haraldsdottir1993minimizing}
Haraldsdottir, J. (1993).
\newblock Minimizing error in the field: Quality control in dietary surveys.
\newblock {\em European Journal of Clinical Nutrition (United Kingdom)}.

\bibitem[Hu et~al., 2018]{hu2018dirichlet}
Hu, J., Reiter, J.~P., Wang, Q., et~al. (2018).
\newblock Dirichlet process mixture models for modeling and generating
  synthetic versions of nested categorical data.
\newblock {\em Bayesian Analysis}, 13(1):183--200.

\bibitem[J{\o}nneland et~al., 1991]{t1991validation}
J{\o}nneland, A., Overvad, K., Haraldsd{\'o}ttir, J., Bang, S., Ewertz, M., and
  Jensen, O.~M. (1991).
\newblock Validation of a semiquantitative food frequency questionnaire
  developed in denmark.
\newblock {\em International Journal of Epidemiology}, 20(4):906--912.

\bibitem[Kant, 2004]{kant2004dietary}
Kant, A.~K. (2004).
\newblock Dietary patterns and health outcomes.
\newblock {\em Journal of the American Dietetic Association}, 104(4):615--635.

\bibitem[Keshteli et~al., 2015]{keshteli2015patterns}
Keshteli, A.~H., Feizi, A., Esmaillzadeh, A., Zaribaf, F., Feinle-Bisset, C.,
  Talley, N.~J., and Adibi, P. (2015).
\newblock Patterns of dietary behaviours identified by latent class analysis
  are associated with chronic uninvestigated dyspepsia.
\newblock {\em British Journal of Nutrition}, 113(05):803--812.

\bibitem[Kipnis et~al., 2009]{kipnis2009modeling}
Kipnis, V., Midthune, D., Buckman, D.~W., Dodd, K.~W., Guenther, P.~M.,
  Krebs-Smith, S.~M., Subar, A.~F., Tooze, J.~A., Carroll, R.~J., and Freedman,
  L.~S. (2009).
\newblock Modeling data with excess zeros and measurement error: Application to
  evaluating relationships between episodically consumed foods and health
  outcomes.
\newblock {\em Biometrics}, 65(4):1003--1010.

\bibitem[Krebs et~al., 1989]{krebs1989ecological}
Krebs, C.~J. et~al. (1989).
\newblock Ecological methodology.
\newblock Technical report, Harper \& Row New York.

\bibitem[Lazarsfeld and Henry, 1968]{lazarsfeld1968latent}
Lazarsfeld, P. and Henry, N. (1968).
\newblock {\em Latent Structure Analysis}.
\newblock Houghton, Mifflin.

\bibitem[Malsiner-Walli et~al., 2016]{malsiner2016model}
Malsiner-Walli, G., Fr{\"u}hwirth-Schnatter, S., and Gr{\"u}n, B. (2016).
\newblock Model-based clustering based on sparse finite gaussian mixtures.
\newblock {\em Statistics and computing}, 26(1-2):303--324.

\bibitem[Medvedovic and Sivaganesan, 2002]{medvedovic2002bayesian}
Medvedovic, M. and Sivaganesan, S. (2002).
\newblock Bayesian infinite mixture model based clustering of gene expression
  profiles.
\newblock {\em Bioinformatics}, 18(9):1194--1206.

\bibitem[Miller and Harrison, 2013]{miller2013simple}
Miller, J.~W. and Harrison, M.~T. (2013).
\newblock A simple example of dirichlet process mixture inconsistency for the
  number of components.
\newblock In {\em Advances in Neural Information Processing Systems}, pages
  199--206.

\bibitem[Miller and Harrison, 2018]{miller2018mixture}
Miller, J.~W. and Harrison, M.~T. (2018).
\newblock Mixture models with a prior on the number of components.
\newblock {\em Journal of the American Statistical Association},
  113(521):340--356.

\bibitem[Motulsky et~al., 1989]{motulsky1989diet}
Motulsky, A.~G. et~al. (1989).
\newblock {\em Diet and Health: Implications for Reducing Chronic Disease
  Risk}.
\newblock National Academies.

\bibitem[Nylund et~al., 2007]{nylund2007deciding}
Nylund, K.~L., Asparouhov, T., and Muth{\'e}n, B.~O. (2007).
\newblock Deciding on the number of classes in latent class analysis and growth
  mixture modeling: A monte carlo simulation study.
\newblock {\em Structural Equation Modeling}, 14(4):535--569.

\bibitem[Papaspiliopoulos and Roberts, 2008]{papaspiliopoulos2008retrospective}
Papaspiliopoulos, O. and Roberts, G.~O. (2008).
\newblock Retrospective markov chain monte carlo methods for dirichlet process
  hierarchical models.
\newblock {\em Biometrika}, 95(1):169--186.

\bibitem[Petrone et~al., 2009]{petrone2009hybrid}
Petrone, S., Guindani, M., and Gelfand, A.~E. (2009).
\newblock Hybrid dirichlet mixture models for functional data.
\newblock {\em Journal of the Royal Statistical Society: Series B},
  71(4):755--782.

\bibitem[Rodriguez et~al., 2008]{rodriguez2012nested}
Rodriguez, A., Dunson, D.~B., and Gelfand, A.~E. (2008).
\newblock The nested dirichlet process.
\newblock {\em Journal of the American Statistical Association},
  103(483):1131--1154.

\bibitem[Rousseau and Mengersen, 2011]{rousseau2011asymptotic}
Rousseau, J. and Mengersen, K. (2011).
\newblock Asymptotic behaviour of the posterior distribution in overfitted
  mixture models.
\newblock {\em Journal of the Royal Statistical Society: Series B},
  73(5):689--710.

\bibitem[Silverwood et~al., 2011]{silverwood2011lca}
Silverwood, R.~J., Nitsch, D., Pierce, M., Kuh, D., and Mishra, G.~D. (2011).
\newblock Characterizing longitudinal patterns of physical activity in
  mid-adulthood using latent class analysis: Results from a prospective cohort
  study.
\newblock {\em American Journal of Epidemiology}, 174(12):1406.

\bibitem[Smith and Khaled, 2012]{smith2012estimation}
Smith, M.~S. and Khaled, M.~A. (2012).
\newblock Estimation of copula models with discrete margins via bayesian data
  augmentation.
\newblock {\em Journal of the American Statistical Association},
  107(497):290--303.

\bibitem[Sotres-Alvarez et~al., 2010]{sotres2010latent}
Sotres-Alvarez, D., Herring, A.~H., and Siega-Riz, A.~M. (2010).
\newblock Latent class analysis is useful to classify pregnant women into
  dietary patterns.
\newblock {\em The Journal of Nutrition}, 140(12):2253--2259.

\bibitem[Sotres-Alvarez et~al., 2013]{sotres2013maternal}
Sotres-Alvarez, D., Siega-Riz, A.~M., Herring, A.~H., Carmichael, S.~L.,
  Feldkamp, M.~L., Hobbs, C.~A., Olshan, A.~F., et~al. (2013).
\newblock Maternal dietary patterns are associated with risk of neural tube and
  congenital heart defects.
\newblock {\em American Journal of Epidemiology}, 177(11):1279--1288.

\bibitem[Subar et~al., 2001]{subar2001comparative}
Subar, A.~F., Thompson, F.~E., Kipnis, V., Midthune, D., Hurwitz, P., McNutt,
  S., McIntosh, A., and Rosenfeld, S. (2001).
\newblock Comparative validation of the block, willett, and national cancer
  institute food frequency questionnaires the eating at america's table study.
\newblock {\em American Journal of Epidemiology}, 154(12):1089--1099.

\bibitem[Teh, 2006]{teh2006hierarchical}
Teh, Y.~W. (2006).
\newblock A hierarchical bayesian language model based on pitman-yor processes.
\newblock In {\em Proceedings of the 21st International Conference on
  Computational Linguistics and the 44th Annual Meeting of the Association for
  Computational Linguistics}, pages 985--992. Association for Computational
  Linguistics.

\bibitem[Teh et~al., 2006]{teh2012hierarchical}
Teh, Y.~W., Jordan, M.~I., Beal, M.~J., and Blei, D.~M. (2006).
\newblock Hierarchical dirichlet processes.
\newblock {\em Journal of the American Statistical Association},
  101(476):1566--1581.

\bibitem[van Havre et~al., 2015]{van2015overfitting}
van Havre, Z., White, N., Rousseau, J., and Mengersen, K. (2015).
\newblock Overfitting bayesian mixture models with an unknown number of
  components.
\newblock {\em PloS one}, 10(7):e0131739.

\bibitem[Venkaiah et~al., 2011]{venkaiah2011application}
Venkaiah, K., Brahmam, G., and Vijayaraghavan, K. (2011).
\newblock Application of factor analysis to identify dietary patterns and use
  of factor scores to study their relationship with nutritional status of adult
  rural populations.
\newblock {\em Journal of health, population, and nutrition}, 29(4):327--338.

\bibitem[Yoon et~al., 2001]{yoon2001national}
Yoon, P.~W., Rasmussen, S.~A., Lynberg, M., Moore, C., Anderka, M., Carmichael,
  S., Costa, P., Druschel, C., Hobbs, C., Romitti, P., et~al. (2001).
\newblock The national birth defects prevention study.
\newblock {\em Public health reports}, 116(1 suppl):32--40.

\bibitem[Zhang et~al., 2004]{zhang2004probabilistic}
Zhang, J., Ghahramani, Z., and Yang, Y. (2004).
\newblock A probabilistic model for online document clustering with application
  to novelty detection.
\newblock {\em Advances in Neural Information Processing Systems},
  4:1617--1624.

\bibitem[Zhang et~al., 2011]{zhang2011new}
Zhang, S., Midthune, D., Guenther, P.~M., Krebs-Smith, S.~M., Kipnis, V., Dodd,
  K.~W., Buckman, D.~W., Tooze, J.~A., Freedman, L., and Carroll, R.~J. (2011).
\newblock A new multivariate measurement error model with zero-inflated dietary
  data, and its application to dietary assessment.
\newblock {\em The Annals of Applied Statistics}, 5(2B):1456--1487.

\end{thebibliography}

\end{document}